\documentclass{JCIN22}
 
\usepackage{amsmath,amsfonts}
\usepackage{algorithmic}
\usepackage{algorithm}
\usepackage{array}
\usepackage{textcomp}
\usepackage{stfloats}
\usepackage{url}
\usepackage{verbatim}
\usepackage{graphicx}
\usepackage{cite}
\usepackage{comment}
\usepackage{breqn,xspace}
\usepackage{color}
\usepackage{booktabs}
\usepackage{tabularx}
\usepackage{gensymb}
\usepackage[numbers,sort&compress]{natbib}

\graphicspath{{fig/}}
\ifodd 1
\else

\fi

\newcommand{\name}{Plotinus\xspace}

\begin{document}

\title{\name: A Satellite Internet Digital Twin System }
%with Modular and Microservice-Oriented Designs

\author[]{Yue~Gao}
\author[]{Kun~Qiu}
\author[]{Zhe~Chen, Wenjun~Zhu, Qi~Zhang, Handong~Luo, Quanwei~Lin, Ziheng~Yang and Wenhao~Liu}

% \author{  \IEEEauthorblockN{Yue~Gao, Kun~Qiu, Zhe~Chen, Wenjun~Zhu, Qi~Zhang, Handong~Luo, Quanwei~Lin, Ziheng~Yang and Wenhao~Liu}
% \thanks{Y. Gao, K. Qiu, Z. Chen, Q. Zhang, H. Luo, Q. Lin, Z. Yang, W. Liu and W. Zhu are with the Intelligent Networking and Computing Research Center and School of Computer Science, Fudan University, Shanghai, China. 
%         (email: gao.yue@fudan.edu.cn; qkun@fudan.edu.cn; zhechen@fudan.edu.cn; qizhang23@m.fudan.edu.cn; hdluo23@m.fudan.edu.cn; qwlin22@m.fudan.edu.cn; zhyang22@m.fudan.edu.cn; liuwh23@m.fudan.edu.cn; wenjun@fudan.edu.cn; ) The correponding author is Kun Qiu.}% <-this % stops a space

% }
        % <-this % stops a space
%\thanks{This paper was produced by the IEEE Publication Technology Group. They are in Piscataway, NJ.}% <-this % stops a space
%\thanks{Manuscript received April 19, 2021; revised August 16, 2021.}}



\maketitle

{\bf \textit{Abstract---}The development of an integrated space-air-ground network (SAGIN) requires sophisticated satellite Internet emulation tools that can handle complex, dynamic topologies and offer in-depth analysis. Existing emulation platforms struggle with challenges like the need for detailed implementation across all network layers, real-time response, and scalability. This paper proposes a digital twin system based on microservices for satellite Internet emulation, namely Plotinus, which aims to solve these problems. Plotinus features a modular design, allowing for easy replacement of the physical layer to emulate different aerial vehicles and analyze channel interference. It also enables replacing path computation methods to simplify testing and deploying algorithms. In particular, Plotinus allows for real-time emulation with live network traffic, enhancing practical network models. The evaluation result shows Plotinus's effective emulation of dynamic satellite networks with real-world devices. Its adaptability for various communication models and algorithm testing highlights Plotinus's role as a vital tool for developing and analyzing SAGIN systems, offering a cross-layer, real-time and scalable digital twin system.
\\[-1.5mm]

}
% \begin{IEEEkeywords}
% Network emulation Framework, 
% Dynamic Topology Adjustment, 
% Microservice Architecture, 
% SAGIN, 
% Real-time Network Adaptability
% \end{IEEEkeywords}

\barefootnote{This work was partially supported by the National Natural Science Foundation of China under Grant 62341105. \\\indent
% Firstname1 Lastname1, Firstname2 Lastname2. Address1, City1 zipcode1, Country1 (e-mail: email1; email2).\\\indent
% Firstname1 Lastname1. Address2, City2 zipcode2, Country2 (e-mail: email1).\\\indent
% Firstname2 Lastname2. Address3, City3 zipcode3, Country3 (e-mail: email2).
Yue Gao, Kun Qiu, Zhe Chen, Wenjun Zhu, Qi Zhang, Handong Luo, Quanwei Lin, Ziheng Yang and Wenhao Liu. The Intelligent Networking and Computing Research Center and School of Computer Science, Fudan University, Shanghai 200433, China (email: gao.yue@fudan.edu.cn; qkun@fudan.edu.cn; zhechen@fudan.edu.cn; wenjun@fudan.edu.cn; qizhang23@m.fudan.edu.cn; hdluo23@m.fudan.edu.cn; qwlin22@m.fudan.edu.cn; zhyang22@m.fudan.edu.cn; liuwh23@m.fudan.edu.cn). The corresponding author is Kun Qiu.
}

\section{Introduction}
Space-air-ground integrated networks (SAGIN), which combine satellites, aerial platforms, and ground infrastructures, are crucial to achieve global connectivity and support advanced satellite Internet communication technologies\upcite{bakareInvestigatingSimulationTechniques}. This integration underscores a transformative shift in communication networks. The convergence of these technologies requires sophisticated emulation platforms to mirror complex interactions within integrated networks, highlighting the need for innovation in this domain\upcite{rayReview6GSpaceairground2022}. Digital twin system\upcite{battyDigitalTwins2018} facilitates real-time monitoring and predictive maintenance, improving decision-making and system performance. As the demand for digital twin systems, e.g., virtual models of a physical world, continues to grow, a digital twin system for satellite networks is negligible.

Digital twin systems play a pivotal role in SAGIN by enabling satellite constellation emulations\upcite{zhaoINTERLINKDigitalTwinAssisted2022}, physical layer analysis, and constellation link calculations. These functionalities address the following critical needs:
a) \textbf{Satellite Constellation Emulation:} Customize the emulation parameters, such as the number of satellites, orbital parameters, communication protocols, etc.
b) \textbf{Physical Layer Analysis:} Accurately model the physical layer characteristics of satellite communications.
c) \textbf{End-to-end Emulation:} Emulate the behavior of upper-layer network protocols and services, including the data link, network, transport, and application layers. 

The increasing complexity of satellite Internet, the growing size of constellations, and the increasing number of inter-satellite and satellite-to-terminal links make digital twin technology indispensable in developing a global communications infrastructure. This situation calls for focused research to utilize digital twin capabilities to enhance SAGIN fully. A satellite Internet digital twin with advanced functionality must embody several key features, such as:

\begin{itemize}
    \item \textbf{Modularized Design:} Supports low-coupled platform modules to easily replace individual modules such as physical layer or routing modules.
    \item \textbf{Flexibility:} Support a variety of aerial vehicles, such as geostationary orbit (GEO), low-orbiting satellites (LEO), and unmanned aerial vehicles (UAV)\upcite{laiStarPerfCharacterizingNetwork2020}.
    \item \textbf{Scalability:} Support the emulation of small-scale to large-scale satellite Internet, facilitating the testing and verifying performance under network conditions of different sizes and complexities.
    \item \textbf{Semi-physical Emulation:} Support integration of real-world devices, with semi-physical emulation capability\textsuperscript{\upcite{liLEOCNRealtimeComplete2024}}.
    \item \textbf{Easy Development}: Support rapid development and deployment with scripting languages.
\end{itemize}

To tackle the aforementioned challenges, we propose an end-to-end satellite Internet digital twin system for SAGIN named Plotinus. Different from the existing emulation platforms such as Hypatia\upcite{kassingExploringInternetSpace2020}, which only supports offline emulation, and the SNS-3 platform\upcite{liFountainCodedStreaming2023}, which only supports GEO satellite, Plotinus has the flexibility to adapt its plugins (e.g., LEO or GEO, path computation algorithms) to accommodate diverse emulation missions. In essence, Plotinus introduces several key advantages to the SAGIN:

\begin{enumerate}
    \item \textbf{End-to-End Emulation:} Support emulation of physical characteristics and network protocol behaviors, including propagation delay, signal attenuation, and behaviors from the data link layer to the application layer.
    \item \textbf{Plugin-based Framework:} Supports various emulation scenarios using different physical layer and path computation plugins.
    \item \textbf{Real-world Integration:} Supports semi-physical emulation of VMs, containers, and physical machines as satellite nodes to enhance the realism of the emulation and test the network in a real-world environment.
\end{enumerate}

The rest of the manuscript is organized as follows: section~\ref{sec:back} introduces the background and related works. In section~\ref{sec:design}, we discuss the comprehensive design of the Plotinus. In section~\ref{sec:feature}, we delve into the modular and real-time features of the Plotinus. Then we assess Plotinus through ping tests and a comparison with the Hypatia platform in section~\ref{sec:evaluation}. We conclude in section~\ref{sec:conclusion}. 

\section{Background and Related Work}
\label{sec:back}
\subsection{Introduction to Satellite Internet Emulation Platforms }

Satellite Internet emulation platform is a key technology in communication technology research. These platforms, indispensable for modeling and evaluating satellite Internet systems, have evolved significantly to meet the increasing complexity of satellite communications\upcite{jiangNetworkSimulatorsSatelliteTerrestrial2023}. Among these, NS-3 is a discrete-event network simulator with extensive capabilities in both wired and wireless domains, including satellite networks\upcite{puttonenSatelliteModelNetwork2014a}. OMNeT++ is recognized for its modular architecture\upcite{niehoeferCNIOpenSource2013}, enabling flexible adaptation to diverse network emulation requirements. satellite tool kit (STK) is renowned for its precise satellite orbit analysis and visualization capabilities\upcite{junqingqiResearchCoverageLink2015a}, proving essential for academic and industrial applications.

\subsection{Diverse Approaches in Satellite Internet Emulations }

Satellite Internet emulation platforms are broadly categorized into three distinct types, each targeting specific facets of satellite Internet analysis:

\begin{itemize}
    \item \textbf{Network Simulators (e.g., NS-3, Hypatia, OS3, STK):} These tools primarily utilize virtual nodes and networks to replicate satellite communication processes\upcite{liuUltraStarLightweightSimulator2023}, offering a combination of theoretical modeling and practical relevance. NS-3 and Hypatia are particularly noted for their intricate satellite network modeling. OMNeT++ and STK contribute comprehensive emulation and analytical capabilities\upcite{bakareInvestigatingSimulationTechniques}.

    \item \textbf{Network Emulators (e.g., Mininet, StarryNet):} Emulators such as Mininet\upcite{tangMiniSaviRealisticSatellite2023} and StarryNet\upcite{laiStarryNetEmpoweringResearchers2023} employ physical or virtual machines to emulate network operations, providing a closer approximation of real-world network behaviors. They are especially beneficial for protocol testing and network configuration in controlled settings.
    
    \item \textbf{Real-world Satellite Testbed:} These platforms provide real-world data by conducting experiments within real-world satellite networks\upcite{panMeasuringLowEarthOrbitSatellite2023}. However, they are constrained by high costs and potential risks from prior experiments and protocol designs.
\end{itemize}

For example, Mininet-Savi\upcite{tangMiniSaviRealisticSatellite2023} is excellent for simulating terrestrial networks, but does not inherently support the unique requirements of satellite communications, such as specialized MAC protocols, physical layer modeling, and accurate channel interference emulation, which are critical for LEO network emulation. 
StarryNet\upcite{laiStarryNetEmpoweringResearchers2023} encounters similar limitations and lacks in-depth modeling of satellite communications at a fine-grained level. The in-depth modeling of the necessary to model the dynamic LEO environment.

On the other hand, SNS-3\upcite{pudduOpenSourceSimulator2022} mainly focuses on GEO networks, providing a powerful platform for these types of satellite systems, but falls short in addressing the high dynamics of LEO constellations. Hypatia\upcite{kassingExploringInternetSpace2020}, although feature-rich, is limited by a flexible algorithmic structure that cannot be easily adapted to the rapidly evolving field of satellite Internet. OS3\upcite{niehoeferCNIOpenSource2013} focuses on orbit simulators and lacks the comprehensive network layer emulation capabilities essential for a comprehensive analysis of satellite communication systems. Therefore, there is an obvious gap in the current consolidated emulation platforms for satellite Internet that can effectively emulate the communication layer of satellite networks, especially for the rapidly changing LEO constellations.

\subsection{The Need for Advanced Digital Twin System}

While each platform offers valuable tools for specific aspects of satellite Internet emulation, none of them provides a complete solution for the accurate modelling of all layers of satellite communication, particularly for the dynamic and complex LEO satellite networks. Recognizing these gaps, there is a clear need for a platform that integrates robust physical and MAC layer modeling, realistic channel interference emulation, packet-level network emulation\upcite{wu2024accelerating}, and AI-based traffic-level analytics\upcite{qiu2022traffic} to represent the complexity of modern satellite communication systems fully.

In summary, the development and diversification of satellite Internet emulation platforms reflect the escalating complexity and requirements of satellite communications research. Considering the limitations of existing tools in capturing the rapid evolution of the satellite Internet environment, the emergence of advanced, realistic, and dynamic digital twin systems is becoming increasingly important for the satellite Internet.

\section{The Design of Plotinus}
\label{sec:design}
\begin{figure*}[!t]
  \centering
  \includegraphics[width=0.8\textwidth]{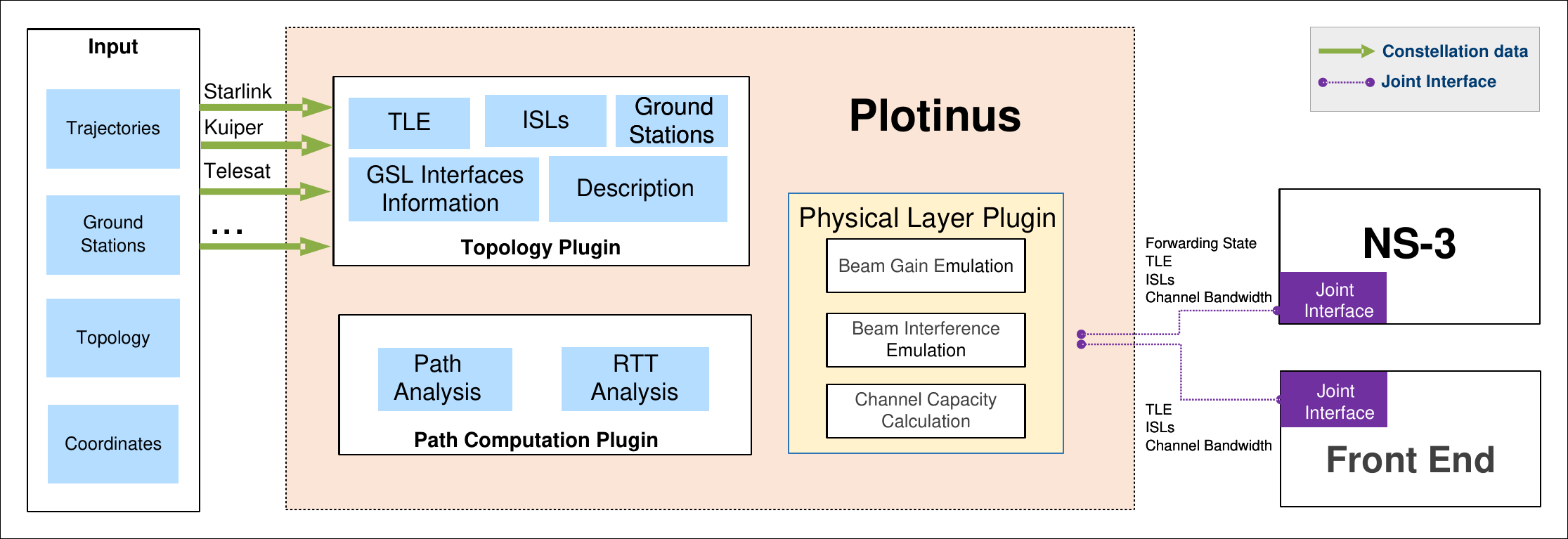}\\
  \caption{An overview architecture of Plotinus.}
  \label{fig:arch}
\end{figure*}

Plotinus is designed to offer a sophisticated emulation environment for analyzing satellite and terrestrial network interactions. The architecture is built on the principle of microservices, ensuring that each component can be individually updated, scaled, and modified without impacting the overall system. The design substantially enhances the platform's utility in various emulation scenarios. Figure~\ref{fig:arch} gives an overview design to show how Plotinus is structured.

\subsection{Microservice Architecture}
Plotinus is designed as a microservice-based architecture. Each plugin is designed as a microservice and is connected using an application program interface~(API). Currently, Plotinus supports physical layer plugins, topology, and path calculation plugins and uses NS-3 as a backend to perform packet-level simulations. Plotinus has designed an interface between these plugins and NS-3 for transferring physical layer parameters and path information to NS-3. In addition, to interact with real-world devices, Plotinus uses TapBridge to connect to any IP-enabled machine, such as virtual machines, containers, commodity servers, or even real satellite nodes.

\subsection{Physical Layer Plugin}
This plugin for Plotinus can emulate the physical layer of satellite communications. This plugin needs to support the emulation of antenna beam behavior, interference between co-frequency beams, and the overall capacity of the communication channel. The plugin calculates how the strength of the antenna beam changes depending on its direction and the satellite's orientation. It needs to accurately calculate the gain of the antenna beam, taking into account the position of the satellite and the specifics of the antenna used. In addition, the plugin evaluates the effect of beam overlap, which can degrade the quality of the transmitted signal. The calculation of channel capacity is an important aspect that the plugin needs to support, considering the quality of the communication link, including how much data can be reliably transmitted over it.

In LEO satellite communications, the state of satellite-to-ground links changes rapidly. We have taken this dynamic change into account in our design. Specifically, the physical layer emulation plugin can calculate the channel capacity of all satellite-to-ground links in the satellite network at different times based on factors such as satellite and ground station locations, antenna information, and noise levels. During the emulation process, Plotinus reads the channel capacity values corresponding to different time slots using the microservice API to represent the status of the satellite-to-ground links. In addition, since satellite-to-ground channel emulation is a complicated task, Plotinus opens the API during the design process. Therefore, any physical layer emulation software may connect with Plotinus through the API, which also demonstrates the feature of Plotinus is easy to extend.

\subsection{Topology and Path Computation Plugin}
These plugins dynamically generate network topologies and compute paths between ground stations or satellites. As satellite positions and the state of the ground station change, the path forwarding table generated by the plugin adapts to these changes, ensuring that the data always selects the most efficient path based on demand. The network path computation is not static; it is updated in real-time to reflect satellite movements and changes, resulting in a highly realistic emulation, which is valuable for planning and analyzing satellite routing and paths. The algorithm in the path computation plugin can be customized to enable path computation for different needs, such as recovery path computation, and can even connect with existing software-defined networks (SDN) controllers with a designed API.

\subsection{Visualisation Interface Plugin}
The visualization interface makes the emulation results more intuitive. It transforms complex network dynamics and routing paths into 3D visual models, making it easier for users to understand and analyze network behavior. The interface not only helps visualize emulation results but also helps identify potential bottlenecks and optimize network configurations. By providing a clear and intuitive view of the network, the plugin significantly increases the usability of Plotinus for network planning and development tasks.

\section{Key features}
\label{sec:feature}
The main features of Plotinus highlight its innovative benefits in satellite Internet emulation. The microservice architecture improves communication efficiency between modules, providing modularity and ease of maintenance. This architecture supports rapid customization and scalability. In addition, its semi-physical emulation capability ensures the timeliness and accuracy of emulation results, providing researchers with a powerful tool to accurately reflect complex network behavior. These capabilities make Plotinus uniquely suited to emulate dynamic, integrated space-air-ground networks, providing a solid foundation for advancing global connectivity and satellite Internet technologies.

\subsection{Microservice Design}
As mentioned, Plotinus uses a microservice design to facilitate communication between plugins. The architecture is built around a comprehensive set of APIs that enable seamless interaction between various plugins. Adopting this design not only simplifies the deployment process and balances the load between different modules but also significantly improves the flexibility and maintainability of the digital twin system.

Each module within Plotinus operates as an independent service through the microservice design, communicating with others via well-defined APIs. This approach offers a modular structure where services can be individually developed, deployed, and scaled. Consequently, it simplifies the updating or replacing of specific components within the simulator. For instance, if the route generation module needs to be upgraded, it can be achieved effortlessly by implementing the required functionalities specified in the API without impacting the rest of the system. Also, this approach provides a modular structure where services can be developed, deployed, and extended individually.  For example, if the path computation needs to be upgraded, only the required functionality specified in the API needs to be implemented without affecting the rest of the system. 

\begin{comment}
Meanwhile, we have implemented the following physical layer model modules for Plotinus expressed in Section IV.D. We replaced the original rudimentary physical layer module with these advanced implementations to enhance the emulation's accuracy and realism in modeling the physical dynamics of satellite communications.

Simultaneously, we develop a dynamic topology generation module presented in Section IV. This module is introduced to replace the existing static topology module. It dynamically adjusts the network topology in response to real-time changes in satellite positions and terrestrial network configurations, ensuring the emulation remains reflective of actual network conditions.

These enhancements underscore our commitment to improving Plotinus's capability to accurately emulate complex satellite and terrestrial network interactions. By integrating these advanced modules, we have significantly expanded the simulator's functionality, offering researchers and network designers a more robust tool for exploring and analyzing network behaviors under various conditions.
\end{comment}

\subsection{Topology Plugin}
The topology plugin in Plotinus is a key component for ensuring accurate emulation of the dynamics of satellites and the configuration of ground networks. The state generation module in the topology plugin generates the satellite network topology based on the relative positions of the satellites within the network. The module utilizes real-time satellite constellation data to adjust the network topology dynamically to reflect the current satellite constellation and ground node configuration. This process allows the digital twin systems to adapt to changes in satellite positions and ensures that the network topology remains up-to-date and reflects the actual situation.

\subsection{Path Computation Plugin}
The path computation plugins work collaboratively with the topology plugin to model and analyze network paths. After the topology plugin generates the network topology, the path computation plugin's path analysis module generates the detailed path analysis data. These data include theoretical round trip time (RTT) calculations, routing paths between nodes, and other key network performance metrics. The path analysis module provides real-time feedback on the most efficient paths between nodes, considering the changing positions of satellites and the terrestrial network environment. The plugin emulates realistic network behavior, including the effect of satellite movement on communication delays and the selection of optimal paths for data transmission.

By integrating the functionality of the topology plugin and the path calculation plugin, Plotinus provides a comprehensive solution for modeling and analyzing the complex dynamics of integrated satellite networks. This approach not only helps to understand the performance characteristics of these networks but also helps to develop and evaluate new network protocols and configurations to address the unique challenges posed by the integration of satellite and terrestrial networks.

\subsection{Physical Layer Plugin}
This section describes the architectural design of the physical layer within the Plotinus, which has been carefully designed to emulate the complex physical characteristics of satellite communications. The plugin consists of three main sub-modules: beam gain emulation, beam interference emulation, and channel capacity calculation.

Since the physical layer plugin can be further developed, all mathematical formulas and algorithms we used here are just for reference. The mathematical formula governing the antenna pattern in Plotinus is expressed as follows\upcite{kimPerformanceAnalysis5G2022}:
\begin{equation}
\begin{aligned}
\mathit{G}_t\left(\theta\right) =
\begin{cases}
\mathit{G}_{\text{max}}, & \text{if } \theta = 0 \\
\mathit4 {G}_{\text{max}}  \left | \frac{J_1(\mathit{k}  \mathit{a}  \sin\theta)}{\mathit{k}  \mathit{a}  \sin\theta} \right |^2, & \text{if } 0 < |\theta| \leq 90^\circ
\end{cases} 
\end{aligned}
\end{equation}

wherein $G_{\text{max}}$ denotes the peak antenna gain, $J_{1}(x)$ represents the Bessel function of the first kind and order, $a$ signifies the antenna's aperture radius, $k=\frac{2\pi f}{c}$ identifies the wave number, with $f$ being the frequency and $c$ the speed of light.

\subsubsection{Beam Gain Emulation}
Precise modeling of the antenna beam gain is critical to any satellite communication system. The beam gain emulation module can simulate an antenna's array gain and calculate the corresponding array gain based on the transmitter position, receiver position, and antenna pattern.

The calculation logic is to determine $\theta$, based on the position of the transmitter and receiver. $\theta$ is the angle measured from the bore sight of the antenna's main beam. Then, the beam gain is calculated based on the antenna pattern.

\subsubsection{Beam Interference Emulation}
The beam interference emulation module within the reference physical layer plugin is dedicated to the precise emulation of inter-beam interference, a critical phenomenon attributed to the superposition of multiple co-frequency beams. This module employs the gain values computed by the beam gain emulation module to quantify the extent of interference. It computes the interference by incorporating the antenna gain pattern and the physical separation between beams. Moreover, it rigorously accounts for the satellite's spatial orientation and positioning to ascertain the interference levels with high fidelity.

The main function of this module is to calculate the interference power between beams sharing the same frequency channel. The interference power, denoted as $P_I$, is formulated as the summation of the power contributions from all beams within the interfering set, mathematically represented as:

\begin{equation}
\begin{aligned}
P_I = \sum_{k \in S} P_{k} |h_{k,j}|^2 
\end{aligned}
\end{equation}

Here, $S$ is the set of transmitters to which the same frequency interference beam belongs, $P_{k}$ signifies the beam transmission power, and $h_{k, j}$ denotes the channel coefficient that encapsulates the channel's characteristics between the $k$th transmitter and the $j$th receiver.

The channel coefficient, $h_{k,j}$, is defined by the equation:

\begin{equation}
\begin{aligned}
h_{k,j} = \frac{\sqrt{G_t(\theta_{k,j}) G_r^j}}{4 \pi \frac{d_{k,j}}{\lambda}} 
\end{aligned}
\end{equation}
where $G_t(\theta_{k,j})$ is the gain of the transmitting antenna from the $k$th transmitter directed towards the $j$th receiver, $G_r^j$ is the gain of the receiving antenna at the $j$th receiver, $d_{k,j}$ denotes the spatial distance between the $k$th transmitter and the $j$th receiver, and $\lambda$ is the wavelength of the propagated signal.

\subsubsection{Channel Capacity Calculation}
The channel capacity calculation module is a critical component of the physical layer plugin, which quantifies the maximum achievable data rate over a satellite link. This rate, known as the channel capacity, encapsulates the upper limit of data throughput without incurring transmission errors. The determination of the channel capacity is rooted in the Shannon theorem, which incorporates the signal-to-interference-plus-noise ratio (SINR), channel bandwidth, and ambient noise levels to produce an accurate measurement.

This module leverages the computed beam gains and interference levels from the beam gain emulation and interference emulation modules to ascertain a precise SINR. Based on SINR and bandwidth values, it calculates the channel capacity.

The Shannon formula computes the channel capacity, expressed as:
\begin{equation}
\begin{aligned}
C_{i,j}^t = Blb(1 + \mathrm{SINR}_{i,j}^t)
\end{aligned}
\end{equation}

where $C_{i,j}^t$ denotes the channel capacity between transmitter $i$ and receiver $j$ at time $t$ in bits per second, $B$ represents the signal bandwidth in Hertz, and $\mathrm{SINR}_{i,j}^t$ is the signal-to-interference-plus-noise ratio for the link between transmitter $i$ and receiver $j$ at time $t$.

The formula for $\mathrm{SINR}_{i,j}^t$ is:

\begin{equation}
\begin{aligned}
\mathrm{SINR}_{i,j}^t = \frac{P_i |h_{i,j}|^2}{k_B T_{\mathrm{rx}} B + \sum_{l \in S} P_l |h_{l,j}|^2} 
\end{aligned}
\end{equation}

In this equation, $P_i |h_{i,j}|^2$ stands for the received power of the desired signals, where $P_i$ is the transmission power and $h_{i,j}$ is the channel coefficient. The term $k_B T_{\mathrm{rx}} B$ denotes the thermal noise power at the receiver, with $k_B$ being the Boltzmann constant, $T_{\mathrm{rx}}$ the receiver's noise temperature, and $B$ the signal bandwidth. The sum $\sum_{l \in S} P_l |h_{l,j}|^2$ represents the accumulated interference power from beams sharing the same frequency.

\subsubsection{Internal Design}

\begin{figure}[!t]
  \centering
  \includegraphics[width=0.3\textwidth]{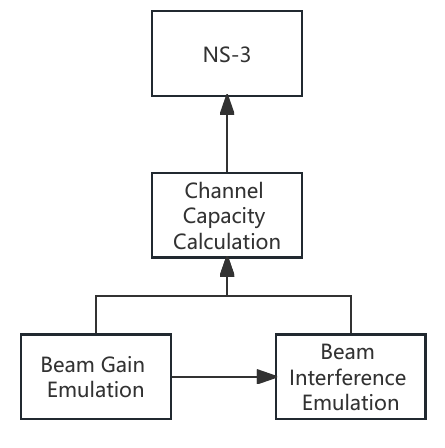}\\
  \caption{The connection between different modules in the reference physical layer plugin.}
  \label{fig:physics module relation}
\end{figure}

The connection between modules is shown in Figure~\ref{fig:physics module relation}. Initially, the beam gain calculation module deduces the antenna's beam gain, while the beam interference calculation module concurrently ascertains the power levels of interference among overlapping beams. Then, the channel capacity calculation module will calculate the channel capacity value based on the input beam gain value and beam interference power.

These calculated channel capacities, pertinent to each satellite-to-ground link within the emulation scenario, are systematically fed into NS-3 for the dynamic emulation process. During the emulation process, NS-3 adjusts the transmission rate based on the channel capacity of the satellite to ground links at different time slots.

\subsection{Real-world Integration}
As a digital twin system, Plotinus is distinguished from other emulators by its support for real-world integration, which enhances emulation by integrating real-world dynamics with the capabilities of NS-3 integrated emulation. Unlike traditional emulators that rely on static network topologies and simulated nodes, Plotinus allows real-time, even interactive (e.g., UI-based) adjustment of the network topology during the emulation. In addition, Plotinus can replace satellite or ground station nodes with any device supporting the IP layer, ensuring a more accurate representation of network behaviors and interactions.

Using NS-3 as a backend, Plotinus creates a virtual environment that closely resembles a real-world network. An important feature of this integration is the ability to expose the emulated network topology to real-world devices. The interface allows real-world devices to connect with the emulated network as if they are part of it, enabling real-time packet processing. In addition, Plotinus can dynamically update paths and network topologies based on changes in satellite positions and terrestrial network configurations. This dynamic update process occurs in real-time, ensuring that the emulation can reflect the current state of the network. As the satellite moves, Plotinus adjusts the path and network parameters accordingly. This feature allows emulation of packet forwarding while considering factors such as propagation delay and link bandwidth, providing insight into network performance and highlighting the platform's ability to emulate complex integrated satellite-terrestrial networks effectively.

\section{Evaluation}
\label{sec:evaluation}
To demonstrate Plotinus's capabilities, we conduct a series of comprehensive experiments. The backend of Plotinus is based on NS-3 (version 3.31), which supports TapBridge. We deploy Plotinus in an Ubuntu 22.04 LTS environment and utilize Linux containers (LXC) to instantiate some ground station nodes to conduct these experiments. This approach emphasizes the platform's modularity and highlights its ability to seamlessly integrate with physical machines, thus demonstrating the high realism of the emulation.

\subsection{Environment}

\begin{figure*}[!t]
  \centering
  \includegraphics[width=0.8\textwidth]{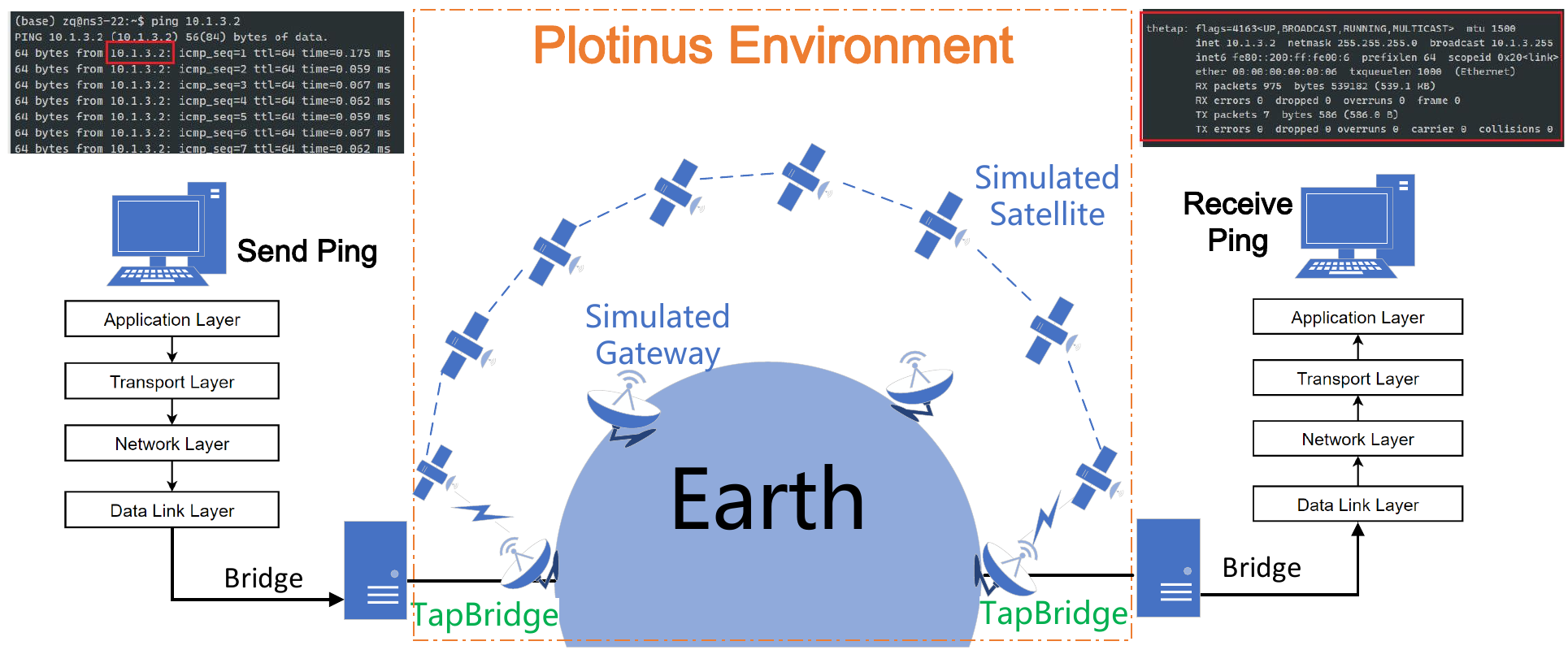}\\
  \caption{The topology of Plotinus evaluation environment using Ping as an example to demonstrate the platform.}
  \label{fig:plt_demo}
\end{figure*}

As shown in Figure~\ref{fig:plt_demo}, the evaluation environment is designed to rigorously test the functionality of Plotinus. The evaluation environment uses connectivity tests (e.g., Ping between devices) to demonstrate the platform's efficacy. We encapsulate the virtual nodes of NS-3 in the LXC, allowing the experiments to be performed on a full network stack. As shown in the figure, the experiment involves sending ping packets from the ground station node on the left to another node on the right. The data arrives at the specified device (left or right ground station node) over a dynamically changing inter-satellite link.

\subsection{Experiment}
In this section, we conduct three sets of experiments to validate the features of Plotinus. TABLE~\ref{tab:table1} lists the parameters of three experiment configurations. All constellation information is obtained from the International Telecommunication Union (ITU)\upcite{itu}. We convert this information to two-line element (TLE) data using some utilities designed in Plotinus. The details of the experiments and results are as follows:

\begin{table}[h!]
  \begin{center}
    \caption{Experiment Configuration}
    \label{tab:table1}
    \begin{tabularx}{\columnwidth}{XXXXX}
        \toprule
        Experi- ment & Constell- ation & Orbits/ Satellites & Apogee (Orbit Altitude) & Inclination Angle  \\
        \midrule
        1) & Kuiper & 34/34 & 630 km & 51.9\degree \\
        2) & Starlink & 72/18 & 550 km & 53.2\degree \\
        3) & Starlink & 72/18 & 550 km & 53.2\degree \\
    
        \bottomrule
    \end{tabularx}
  \end{center}
\end{table}

\subsubsection{Satellite-to-Ground Link}
This experiment by Plotinus is designed to test the effect of changes in physical layer channel capacity on transmission control protocol (TCP) send rates. The channel capacity refers to the maximum transmission rate that can be achieved for error-free transmission over a wireless channel. The channel capacity limits the TCP send rate. To verify this, experiments are performed under two conditions: a stable channel capacity and a rapidly changing channel capacity. End-to-end communication emulations are performed in each case, and the change in TCP send rate is recorded and analyzed.

The Kuiper constellation is chosen as the emulation scenario for the experiment. The Kuiper constellation consists of $34$ orbital planes, with $34$ satellites per plane and an orbital altitude of $630$ km. The experiment uses relay transmission only at the ground stations, and the end-to-end communication is carried out between one ground station in London and another in Shanghai. The emulation time slot was set to $1$ s, and the total emulation duration was $200$ s. The channel capacity is set to $10$ Mbit/s for a stable channel and varies between $1$ Mbit/s and $10$ Mbit/s for a rapidly changing channel.

\subsubsection{Path Computation}
In Plotinus, dynamic changes in satellite network topology are simulated, and customized path-computing algorithms are incorporated into the path-computing module to optimise satellite network paths. Plotinus provides a satellite state awareness algorithm, which enables each satellite to communicate with surrounding satellites through inter-satellite links dynamically and independently decides the direction of path forwarding based on the acquired information. This experiment highlights the advantages of Plotinus's modular design by comparing it with the Hypatia platform. It provides a solution for the dynamic changes in satellite networks and validates its effectiveness.

The experiment compares the Hypatia platform with the Plotinus platform using the Starlink constellation as the emulation scenario. The Starlink constellation consists of $72$ orbital planes, with $18$ satellites in each plane. The satellite orbit of the Starlink constellation is chosen in this experiment of $53$\degree. In Hypatia, all inter-satellite links are assumed to be operational, as Hypatia does not support setting up failed nodes or inter-satellite links. It uses a centralized shortest-path computing algorithm by default. In Plotinus, $10\%$ of the inter-satellite links are set to fail, and the satellite state awareness algorithm is used for path computing. The emulation time slots are set to $100$ ms, and the total emulation time is $200$ s.

\subsubsection{Real-world Integration}
Plotinus can support different constellation configurations and ground stations. The Starlink constellation is chosen in this experiment, and the ground stations are located in Shanghai and São Paulo. The emulation duration is set to $200$ s. We integrate two Linux containers (LXC) into two ground stations and send Ping packets from Shanghai to São Paulo, and vice versa.

\begin{comment}
Plotinus can simulate satellite network topology changes and use customized path computation algorithms in the path computation plugin to optimize paths in satellite network. Plotinus supports the emulation of inter-satellite information. Each satellite can communicate with the surrounding satellites through inter-satellite links and independently decide the path forwarding direction based on the acquired information. This experiment highlights the advantages of the modular design of Plotinus by comparing it with the Hypatia platform. It provides a solution for dynamic changes in satellite networks and validates its effectiveness.
\end{comment}

\subsection{Results and Analysis}

\subsubsection{Satellite-to-Ground Link}
\begin{figure}[!t]
  \centering
  \includegraphics[width=0.4\textwidth]{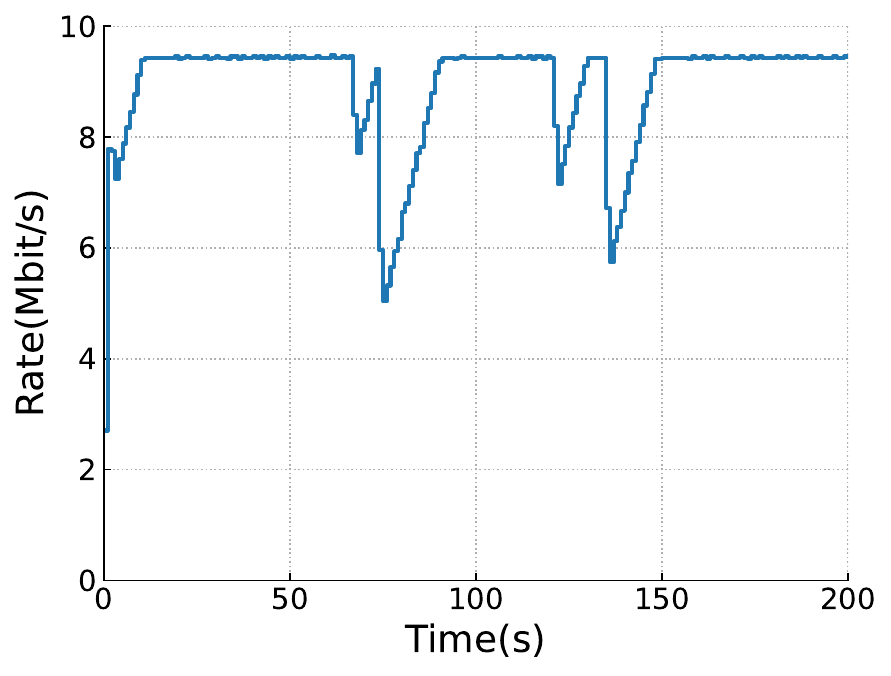}\\
  \caption{TCP sending rate for a stable channel.}
  \label{fig:stable_tcp}
\end{figure}

\begin{figure}[!t]
  \centering
  \includegraphics[width=0.4\textwidth]{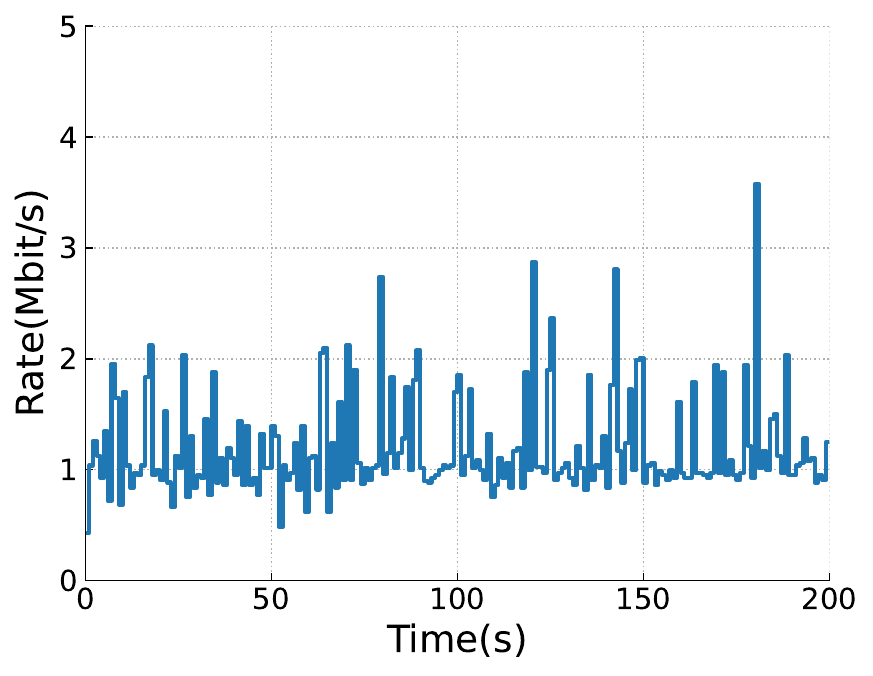}
  \caption{TCP sending rate for a rapidly changing channel.}
  \label{fig:drastical_tcp}
\end{figure}

As shown in Figure~\ref{fig:stable_tcp}, the TCP sending rate remains almost constant when the channel is stable. However, as shown in Figure~\ref{fig:drastical_tcp}, the TCP sending rate also fluctuates significantly when the channel rapidly changes. This demonstrates that physical layer channel capacity variations can affect Plotinus's TCP sending rate.

\subsubsection{Path Computation}
\begin{figure}[!t]
  \centering
  \includegraphics[width=0.4\textwidth]{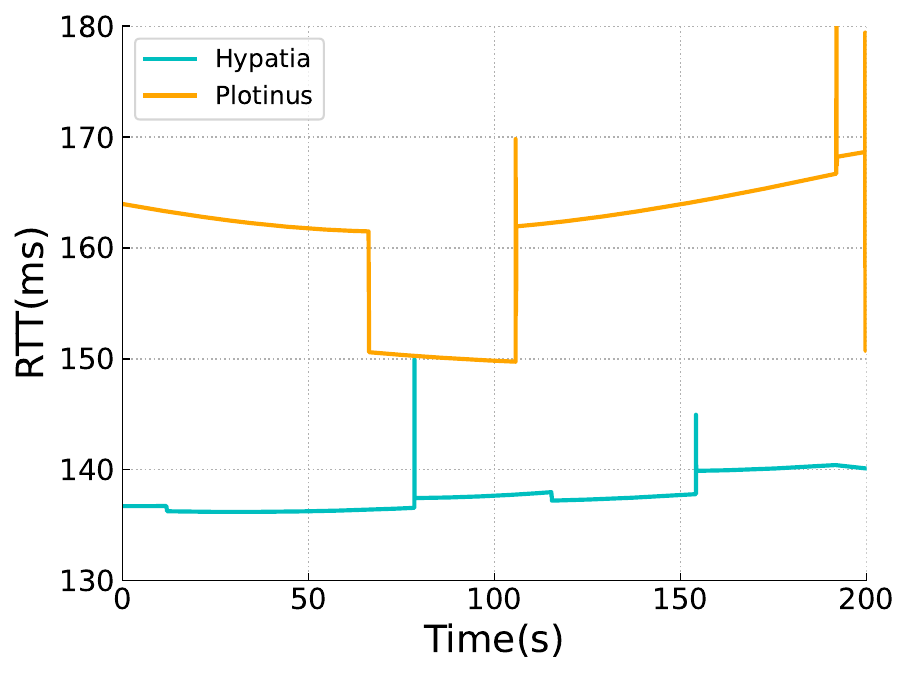}
  \caption{The comparison of RTT between Hypatia and Plotinus.}
  \label{fig:compare_rtt}
\end{figure}

As shown in Figure~\ref{fig:compare_rtt}, the network performance of Plotinus is slightly inferior to that of Hypatia. This is because Plotinus has a $10\%$ failure rate for inter-satellite links, which can result in non-optimal paths when computing forwarding paths. On the other hand, Hypatia does not have any failed inter-satellite links, so all path computing algorithms are based on the shortest route. This experiment demonstrates that Plotinus has robust scalability and can support different path computing methods. The architectural design of Plotinus enables it to leverage its advantages when facing various challenges in satellite networks.

\begin{figure*}[!htp]
  \centering
  \includegraphics[width=0.9\textwidth]{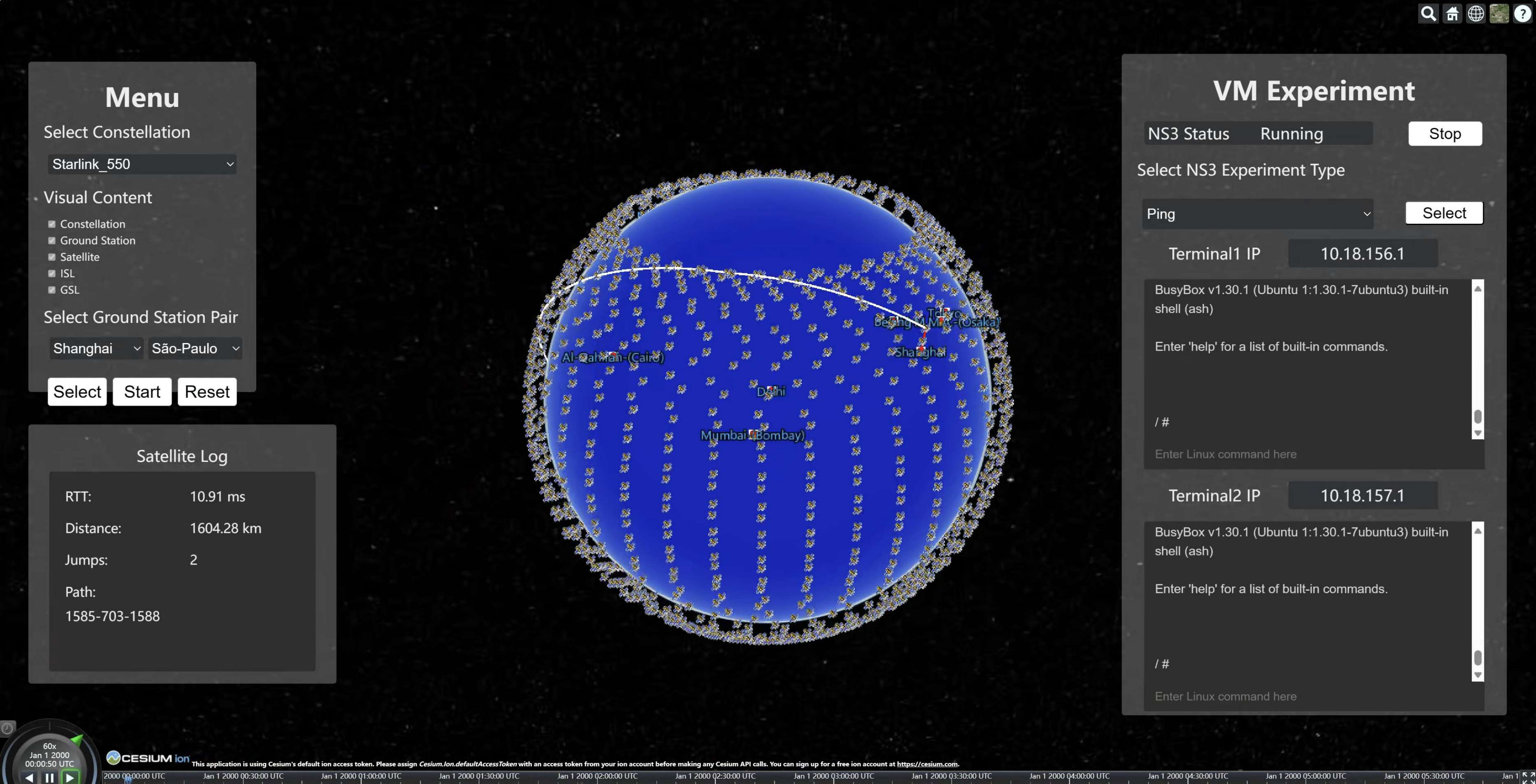}
  \caption{The front end of Plotinus displaying network performance metrics such as RTT, distance, jumps and paths.}
  \label{fig:plotinus_frontend}
\end{figure*}

\subsubsection{Real-world Integration}
As shown in (Figure~\ref{fig:plotinus_frontend}), the left section of Plotinus's front end allows for selecting constellations and ground stations for end-to-end emulation. It also displays network performance metrics such as RTT, distance, hops, and paths. The right section represents the real-time end-to-end communication module, enabling interactive communication using the ping command at different times.

\begin{figure}[!htp]
  \centering
  \includegraphics[width=0.4\textwidth]{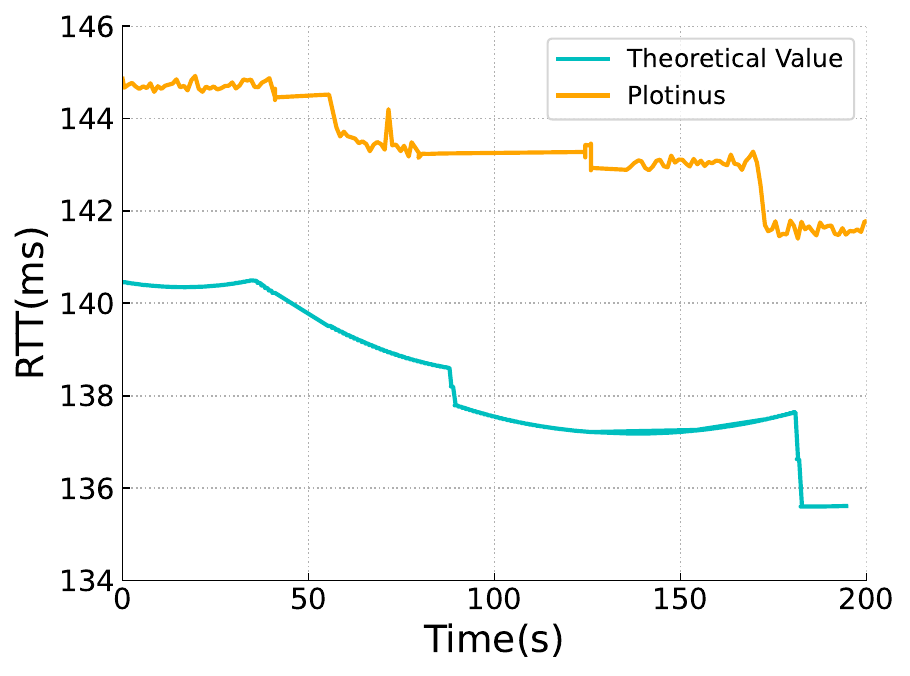}
  \caption{RTT between Shanghai and São Paulo.}
  \label{fig:rtt_200s}
\end{figure}

Figure~\ref{fig:rtt_200s} shows the variation of the RTT within the $200$-second experiment. The green line represents the theoretical value of the end-to-end communication latency. This result is obtained by modeling the satellite network, obtaining the network topology, and calculating the changing propagation delay over time, resulting in a relatively smooth curve. The orange line represents the real-time end-to-end communication delay obtained using Plotinus. Plotinus emulates the packet transmission process from the sender to the receiver and can emulate container processing delays, making the experiment more realistic. Therefore, the results obtained through Plotinus are slightly higher than the theoretical values. Moreover, extra latency is caused by the overhead between Tapbridge and Linux Container. The result indicates that Plotinus can effectively emulate end-to-end communication.

\section{Conclusion}
\label{sec:conclusion}
In this manuscript, we introduce Plotinus, an innovative digital twin system tailored for satellite Internet within the space-air-ground integrated network. Plotinus leverages a microservice-based structure to tackle challenges, including scalability, immediate responsiveness, and a modular approach. Through rigorous testing, Plotinus has proven its ability to replicate dynamic network environments and evaluate path computation strategies effectively. The core features of Plotinus, such as modular design, scalability, support for real-world integration, and user-friendly development process, align with the critical necessities for cutting-edge digital twin technologies. This significantly simplifies the complexities involved in designing satellite Internet. Plotinus provides a solid foundation for future exploration and improvement of satellite Internet. It highlights the system's ability to lead innovation and enhance the development and fine-tuning of communications infrastructure.

% \section*{Acknowledgements}
% This work was partially supported by the National Natural Science Foundation of China under Grant $62341105$.

\bibliography{reference}

\begin{thebibliography}{10}
\providecommand{\url}[1]{#1}
\csname url@samestyle\endcsname
\providecommand{\newblock}{\relax}
\providecommand{\bibinfo}[2]{#2}
\providecommand{\BIBentrySTDinterwordspacing}{\spaceskip=0pt\relax}
\providecommand{\BIBentryALTinterwordstretchfactor}{4}
\providecommand{\BIBentryALTinterwordspacing}{\spaceskip=\fontdimen2\font plus
\BIBentryALTinterwordstretchfactor\fontdimen3\font minus \fontdimen4\font\relax}
\providecommand{\BIBforeignlanguage}[2]{{%
\expandafter\ifx\csname l@#1\endcsname\relax
\typeout{** WARNING: IEEEtran.bst: No hyphenation pattern has been}%
\typeout{** loaded for the language `#1'. Using the pattern for}%
\typeout{** the default language instead.}%
\else
\language=\csname l@#1\endcsname
\fi
#2}}
\providecommand{\BIBdecl}{\relax}
\BIBdecl

\bibitem{bakareInvestigatingSimulationTechniques}
BAKARE B I, ENOCH J D. Investigating Some Simulation Techniques for Wireless Communication System[J]. IOSR Journal of Electronics and Communication Engineering, 2019, 14(3): 56-65.

\bibitem{rayReview6GSpaceairground2022}
RAY P P. A Review on 6G for Space-Air-Ground Integrated Network: Key Enablers, Open Challenges, and Future Direction[J]. Journal of King Saud University - Computer and Information Sciences, 2022, 34(9): 6949-6976.

\bibitem{battyDigitalTwins2018}
BATTY M. Digital Twins[J]. Environment and Planning B: Urban Analytics and City Science, 2018, 45(5): 817-820.

\bibitem{zhaoINTERLINKDigitalTwinAssisted2022}
ZHAO L, WANG C, ZHAO K, et al. INTERLINK: A Digital Twin-Assisted Storage Strategy for Satellite-Terrestrial Networks[J]. IEEE Transactions on Aerospace and Electronic Systems, 2022, 58(5): 3746-3759.

\bibitem{laiStarPerfCharacterizingNetwork2020}
LAI Z, LI H, LI J. StarPerf: Characterizing Network Performance for Emerging Mega-Constellations[C].//2020 IEEE 28th International Conference on Network Protocols. Piscataway: IEEE Press, 2020: 1-11.

\bibitem{liLEOCNRealtimeComplete2024}
LI C, WANG H, LIANG W, et al. LEOCN: Real-time and complete network simulation framework for LEO constellation networks[J]. Journal of High Speed Networks, 2024, 30(1): 1-18.

\bibitem{kassingExploringInternetSpace2020}
KASSING S, BHATTACHERJEE D, {\'A}GUAS A B, et al. Exploring the "Internet from Space" with Hypatia[C].//Proceedings of the ACM Internet Measurement Conference. New York: ACM Press, 2020: 214-229.

\bibitem{liFountainCodedStreaming2023}
LI Y, FENG R, GAO R, et al. Fountain Coded Streaming for SAGIN With Learning-Based Pause-and-Listen[J]. IEEE Networking Letters, 2023, 5(1): 36-40.

\bibitem{jiangNetworkSimulatorsSatelliteTerrestrial2023}
JIANG W, ZHAN Y, XIAO X, et al. Network Simulators for Satellite-Terrestrial Integrated Networks: A Survey[J]. IEEE Access, 2023, 11: 98268-98262.

\bibitem{puttonenSatelliteModelNetwork2014a}
PUTTONEN J, RANTANEN S, LAAKSO F, et al. Satellite model for network simulator 3[C].//Proceedings of the 7th International ICST Conference on Simulation Tools and Techniques. Brussels: ICST Press, 2014: 86-91.

\bibitem{niehoeferCNIOpenSource2013}
NIEHOEFER B, \v{S}UBIK S, WIETFELD C. The CNI open source satellite simulator based on OMNeT++[C].//Proceedings of the 6th International ICST Conference on Simulation Tools and Techniques. Brussels: ICST Press, 2013: 314-321.

\bibitem{junqingqiResearchCoverageLink2015a}
QI J, LI Z, LIU G. Research on Coverage and Link of Multi-Layer Satellite Network Based on STK[C].//2015 10th International Conference on Communications and Networking in China. Piscataway: IEEE Press, 2015: 410-415.

\bibitem{liuUltraStarLightweightSimulator2023}
LIU X, MA T, TANG Z, et al. UltraStar: A Lightweight Simulator of Ultra-Dense LEO Satellite Constellation Networking for 6G[J]. IEEE/CAA Journal of Automatica Sinica, 2023, 10(3): 632-645.

\bibitem{tangMiniSaviRealisticSatellite2023}
TANG Z, ZHAO J, LI H, et al. Mini-Savi: Realistic Satellite Network Simulation Platform Based on Open-Source Tools[C].//2023 Fourth International Conference on Frontiers of Computers and Communication Engineering. Piscataway: IEEE Press, 2023: 27-30.

\bibitem{laiStarryNetEmpoweringResearchers2023}
LAI Z, LI H, DENG Y, et al. StarryNet: Empowering Researchers to Evaluate Futuristic Integrated Space and Terrestrial Networks[C].//20th USENIX Symposium on Networked Systems Design and Implementation. Berkeley: USENIX Press, 2023: 1309-1324.

\bibitem{panMeasuringLowEarthOrbitSatellite2023}
PAN J, ZHAO J, CAI L. Measuring a Low-Earth-Orbit Satellite Network[C].//2023 IEEE 34th Annual International Symposium on Personal, Indoor and Mobile Radio Communications. Piscataway: IEEE Press, 2023: 1-6.

\bibitem{pudduOpenSourceSimulator2022}
PUDDU R, POPESCU V, MURRONI M. An Open Source Simulator for Next Generation Satellite Broadband Traffic Management[C].//2022 IEEE Aerospace Conference. Piscataway: IEEE Press, 2022: 1-6.

\bibitem{wu2024accelerating}
WU J, SU S, WANG X, et al. Accelerating Handover in Mobile Satellite Network[C].//IEEE International Conference on Computer Communications. Piscataway: IEEE Press, 2024.

\bibitem{qiu2022traffic}
QIU K, CHANG H, WANG Y, et al. Traffic Analytics Development Kits (TADK): Enable Real-Time AI Inference in Networking Apps[C].//IEEE International Conference on Ubiquitous and Future Networks. Piscataway: IEEE Press, 2022.

\bibitem{kimPerformanceAnalysis5G2022}
KIM P, RYU J G, PARK S. Performance Analysis for 5G/6G Satellite Communication under Nonlinear HPA Channel[C].//2022 27th Asia Pacific Conference on Communications. Piscataway: IEEE Press, 2022: 641-642.

\bibitem{itu}
ITU Space Network Systems Online[DB/OL]. \url{https://www.itu.int/sns/.}

\end{thebibliography}
\bibliographystyle{IEEEtran}

\section*{About the Authors}\footnotesize\vskip 2mm

\parpic{\includegraphics[width=22mm,height=30mm]{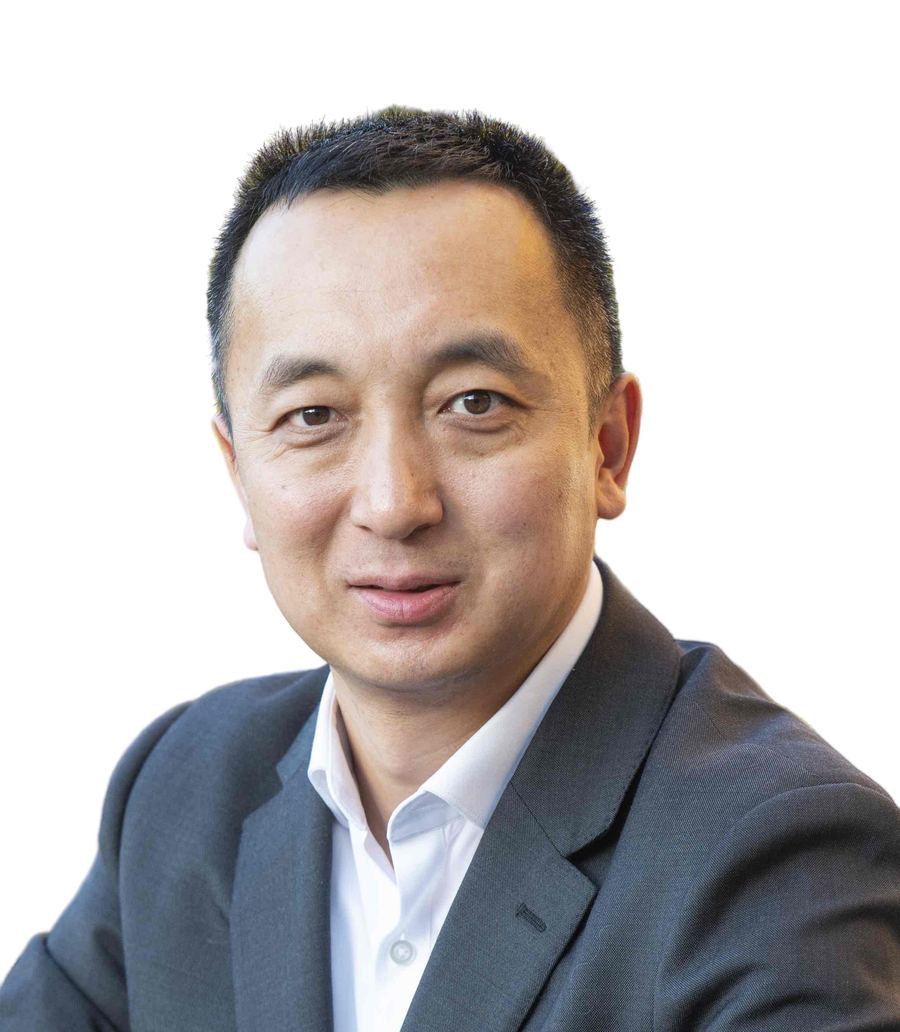}}%
\noindent{\bf Yue Gao} received his PhD from the Queen Mary University of London, UK, in 2007. He is a Chair Professor at the School of Computer Science, Director of the Intelligent Networking and Computing Research Centre at Fudan University, China and a Visiting Professor at the University of Surrey, UK. His research interests include smart antennas, sparse signal processing and cognitive networks for mobile and satellite systems. He is a Fellow of the IEEE.
\vskip4.59mm

\parpic{\includegraphics[width=22mm,height=30mm]{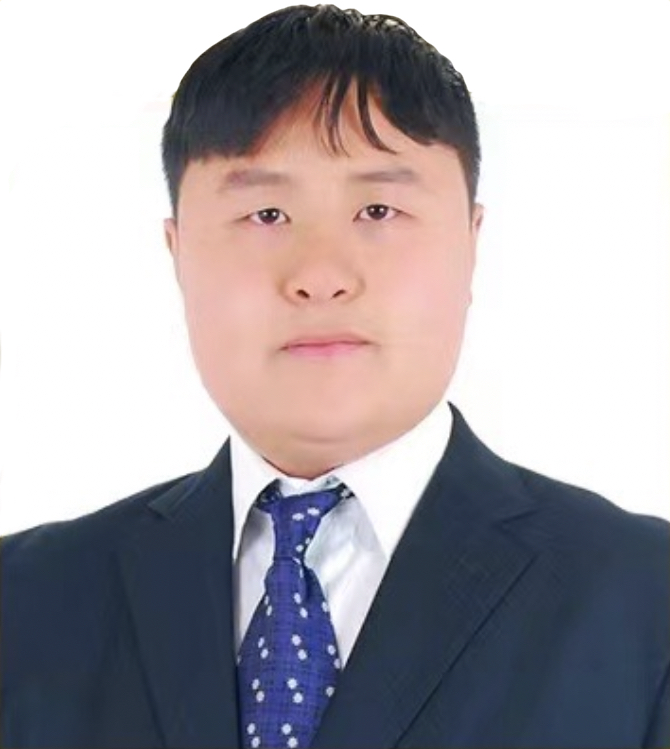}}%
\noindent{\bf Kun Qiu} received his B.Sc. from Fudan University in 2013 and his PhD from Fudan University in 2019. He works for Intel as a software engineer from 2019 to 2023. He joined Fudan University in 2023 as an Assistant Professor in the School of Computer Science at Fudan University. His research interests include computer networks and computer architecture. He is a member of IEEE, ACM, and CCF.
\vskip4.59mm

\parpic{\includegraphics[width=22mm,height=30mm]{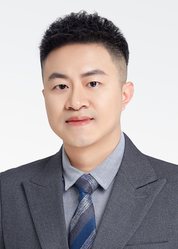}}%
\noindent{\bf Zhe Chen} received his PhD in Computer Science from Fudan University, China, with a 2019 ACM SIGCOMM China Doctoral Dissertation Award. He is an Assistant Professor in the School of Computer Science at Fudan University and the Co-Founder of AIWiSe Ltd. Inc. Before joining Fudan University, he worked as a research fellow at NTU for three years, and his research achievements, along with his efforts in launching products based on them, have thus earned him 2021 ACM SIGMOBILE China Rising Star Award recently.
\vskip4.59mm

\parpic{\includegraphics[width=22mm,height=30mm]{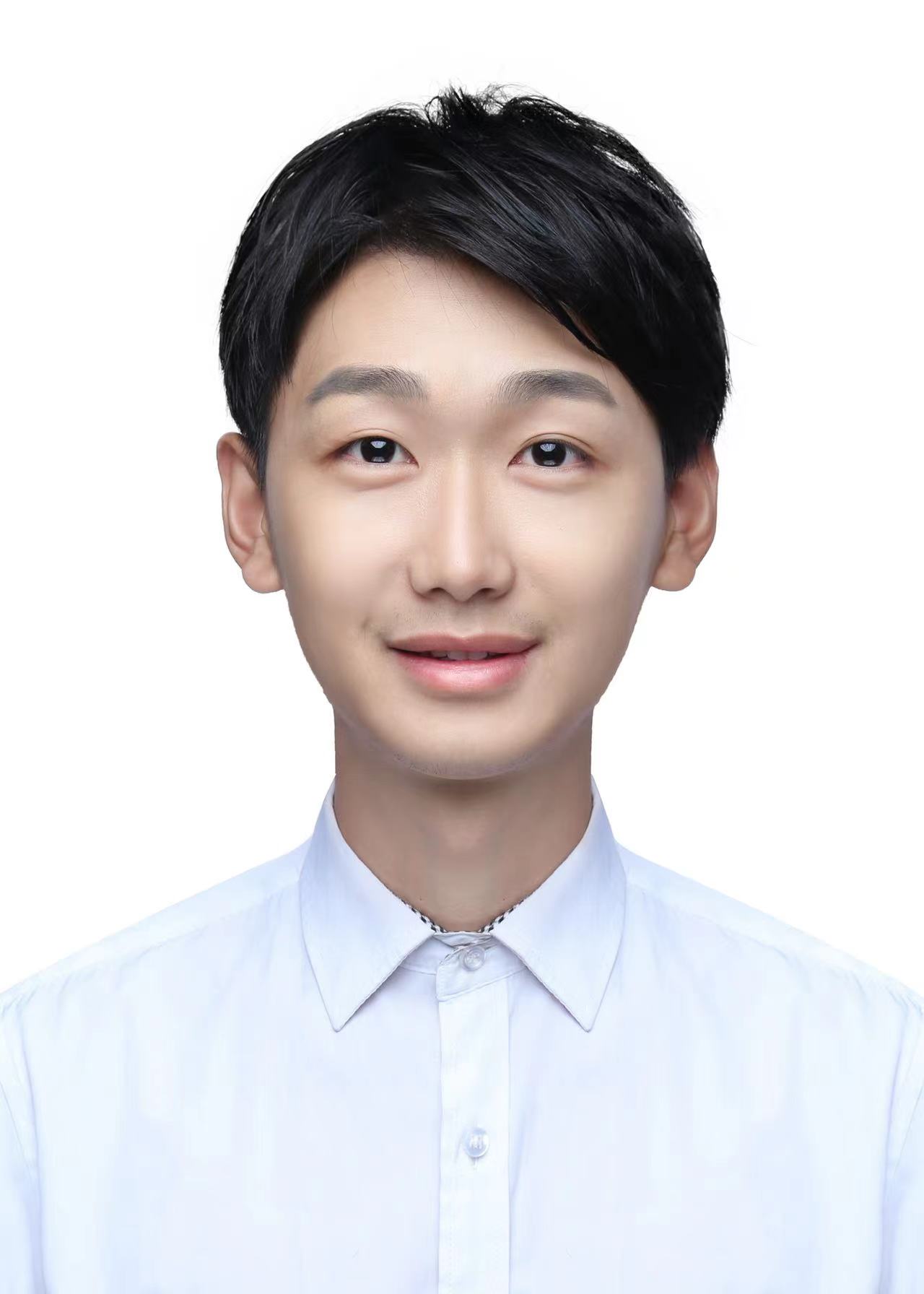}}%
\noindent{\bf Wenjun Zhu} received his master's degree from  Nanjing University of Posts and Telecommunications in 2020. He works for Intel as a software engineer from 2020 to 2023. Now, he joined Fudan University in 2023 as a software engineer. His research interests include computer networks and computer architecture.
\vskip4.59mm

\parpic{\includegraphics[width=22mm,height=30mm]{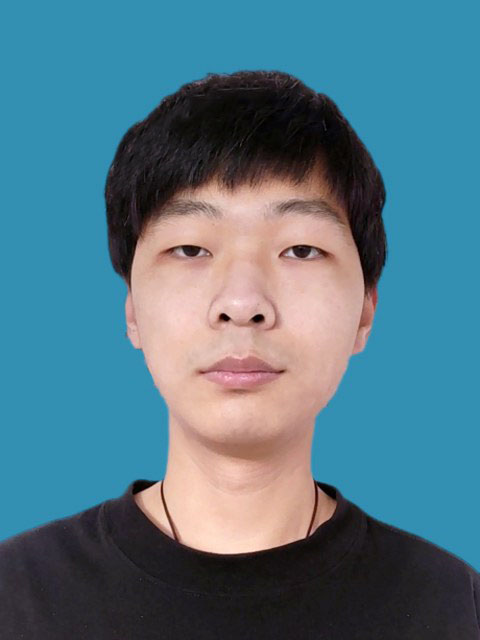}}%
\noindent{\bf Qi Zhang} earned his B.Eng. degree from Henan University in 2020 and his M.Eng. in Computer Science and Technology from Soochow University in 2023. He is currently pursuing his PhD at Fudan University. Zhang has made significant contributions to the field of edge computing. His research primarily focuses on routing optimization in integrated space-air-ground networks, a cutting-edge area with vast potential applications.
\vskip4.59mm

\parpic{\includegraphics[width=22mm,height=30mm]{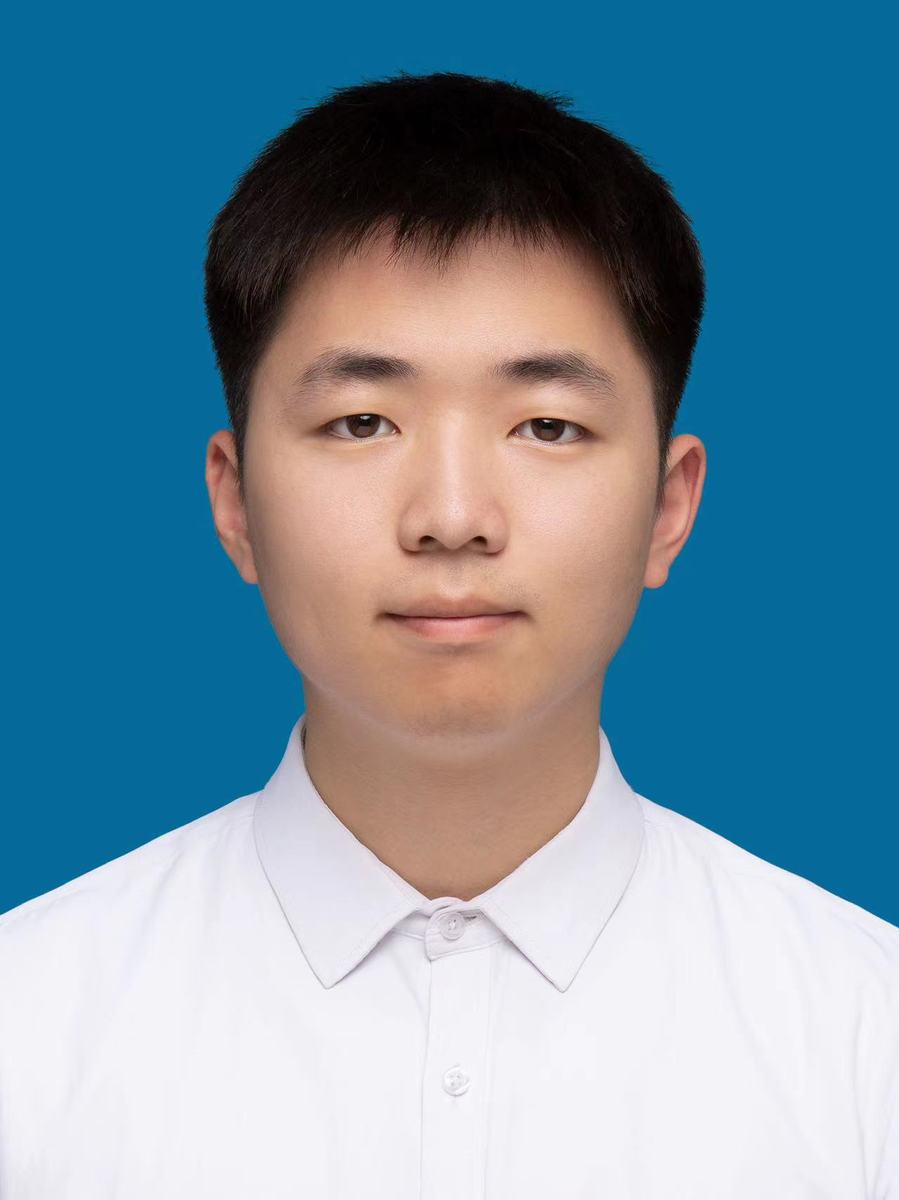}}%
\noindent{\bf Handong Luo} received his bachelor's degree in Computer Science from Hangzhou Dianzi University, Hangzhou, China, in 2023. He is working toward a master's degree with the School of Computer Science, Fudan University, Shanghai, China. His research interests include satellite network routing and scalability.
\vskip4.59mm

\parpic{\includegraphics[width=22mm,height=30mm]{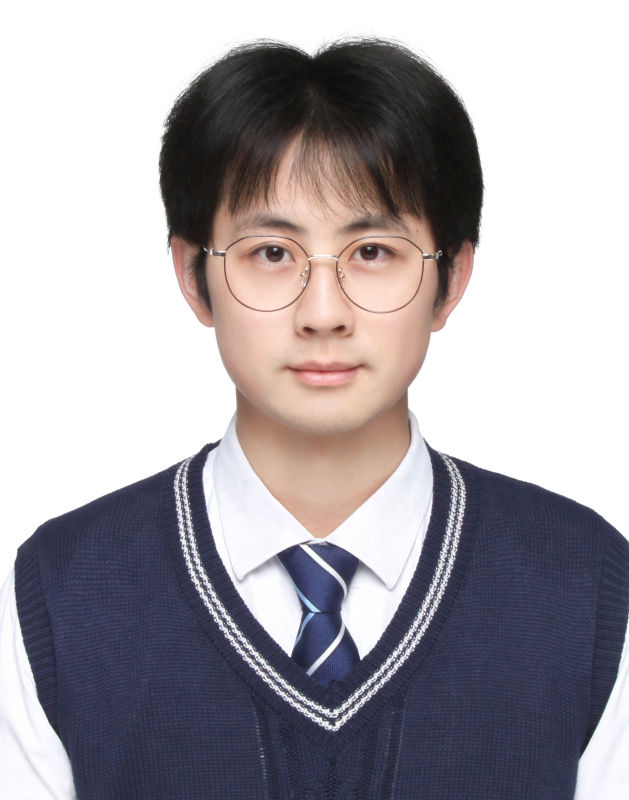}}%
\noindent{\bf Quanwei Lin} received his bachelor's degree in Engineering Management from Nanjing Audit University, Nanjing, China, in 2021. He is working toward a master's degree with the School of Computer Science, Fudan University, Shanghai, China. His research interests include satellite network routing and link quality management.
\vskip4.59mm

\parpic{\includegraphics[width=22mm,height=30mm]{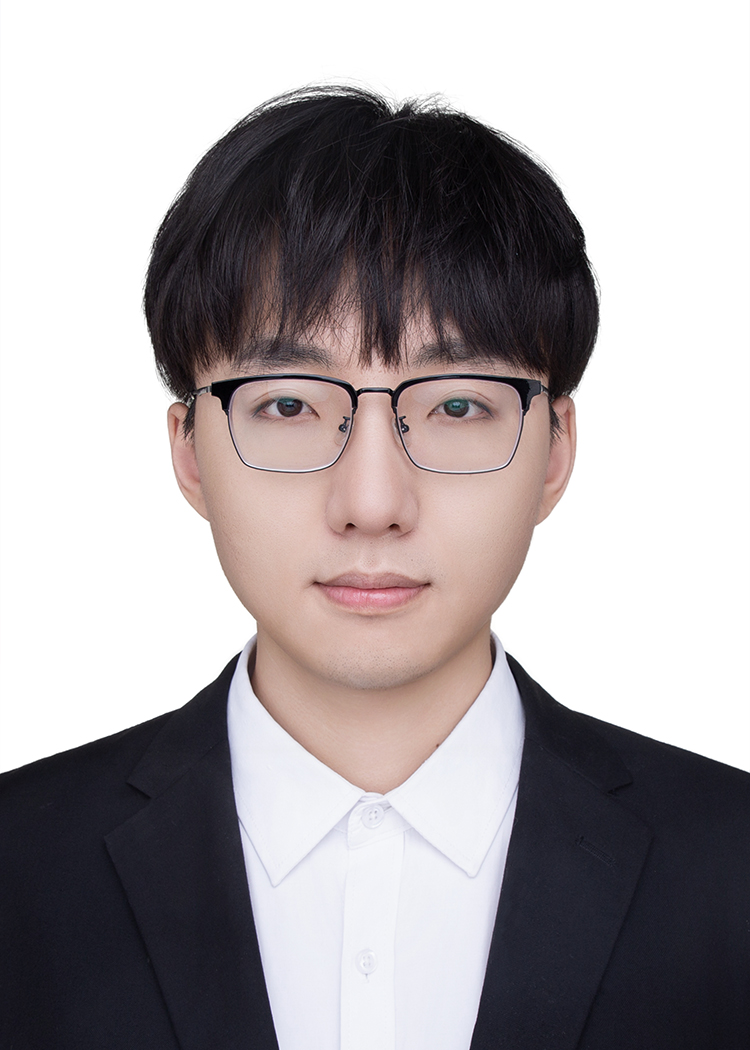}}%
\noindent{\bf Ziheng Yang} earned his bachelor’s degree in communication engineering from Nanjing University of Posts and Telecommunications in 2022. He is pursuing his master's degree at the School of Computer Science, Fudan University, Shanghai, China. His research interests include satellite communication and beam hopping.
\vskip4.59mm

\parpic{\includegraphics[width=22mm,height=30mm]{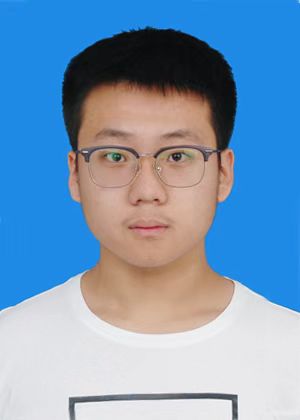}}%
\noindent{\bf Wenhao Liu} received his bachelor’s degree in Computer Science from Fudan University, Shanghai, China, in 2023. He is working toward a PhD at the School of Computer Science, Fudan University, Shanghai, China. His research interests include satellite network scalability and resilient architecture.

\vfill

\end{document}